\documentclass[aps,prb,twocolumn, groupedaddress,showpacs,floatfix]{revtex4}
\usepackage[centertags]{amsmath}
\usepackage{graphicx,amssymb}

\begin{document}
\title{Competition between band and Mott insulator in the
bilayer Hubbard model: a dynamical cluster approximation study}

\author{Hunpyo Lee$^{2}$}
\email[]{hplee@itp.uni-frankfurt.de}
\author{Yu-Zhong Zhang$^1$}
\email[]{yzzhang@tongji.edu.cn}
\author{Harald O. Jeschke$^2$}
\author{Roser Valent\'\i$^2$}
\affiliation{$^1$Shanghai Key Laboratory of Special Artificial Microstructure Materials and Technology, \\
School of Physics Science and Engineering, Tongji University, Shanghai 200092, P.R. China}
\affiliation{$^2$Institut f\"ur Theoretische Physik, Goethe-Universit\"at
Frankfurt, Max-von-Laue-Stra{\ss}e 1, 60438 Frankfurt am Main, Germany}
\date{\today}

\begin{abstract}
  We investigate the nature of the insulating phases in a bilayer
  Hubbard model with intralayer coupling $t$ and interlayer coupling
  $t_{\perp}$ at large interaction strength $U/t$ and half-filling. We
  consider a dynamical cluster approximation with a cluster size of
  $N_c=2\times4$, where short-range spatial fluctuations as well as
  on-site dynamical fluctuations are emphasized.  By varying the band
  splitting ($t_{\perp}/t$), we find that at $t_{\perp}/t\simeq1.5$
  the Mott behavior is rapidly suppressed in the momentum sectors
  ($\pi,0$) and ($0,\pi$).  At $t_{\perp}/t\simeq2.5$ Mott features
  dominate in the momentum sectors ($\pi,\pi$) of the bonding band and
  ($0,0$) of the anti-bonding band and at $t_{\perp}/t\simeq3.0$ a
  tiny scattering rate is observed in all momentum sectors at the
  Fermi level, indicating a transition from a Mott to a band
  insulator.  We attribute such a momentum-dependent evolution of the
  insulating behavior to the competition and cooperation between
  short-range spatial fluctuations and interlayer coupling $t_{\perp}$
  with the help of the Coulomb interaction $U$. Finally, we also
  discuss the possible appearance of non-Fermi liquid behavior away
  from half-filling.
\end{abstract}

\maketitle

\section{Introduction\label{Introduction}}
During the last decade intensive debates have centered around the
question what happens when a system evolves from a band to a Mott
insulator~\cite{Mott1949,Gebhard} in the context of different models
in different
dimensions~\cite{Nakamura2000,Zhang2004,Tam2006,Ejima2007,Fabrizio1999,
Zhang2003,Batista2004,Go2011,Garg2006,Kancharla2007PRL,Paris2007,Craco2008,Chen2010}.
like the extended
 Hubbard model~\cite{Nakamura2000,Zhang2004,Tam2006,Ejima2007}
and the ionic
Hubbard model~\cite{Fabrizio1999,Zhang2003,Batista2004,Go2011} in one
dimension, or
the ionic Hubbard model~\cite{Garg2006,Kancharla2007PRL,Paris2007,Chen2010}
in two dimensions.  A
model that has gained a lot of attention in recent years is the
bilayer Hubbard model on a square lattice.  The discovery of bilayer
band splitting in angle-resolved photoemission spectroscopy
experiments~\cite{Damascelli2003} for YBa$_2$Cu$_3$O$_{6+x}$
(YBCO)~\cite{Fournier2010} suggested the consideration of such a model
as a minimal model for describing double-layered YBCO compounds.

\begin{figure}[htbp]
\includegraphics[width=0.4\textwidth]{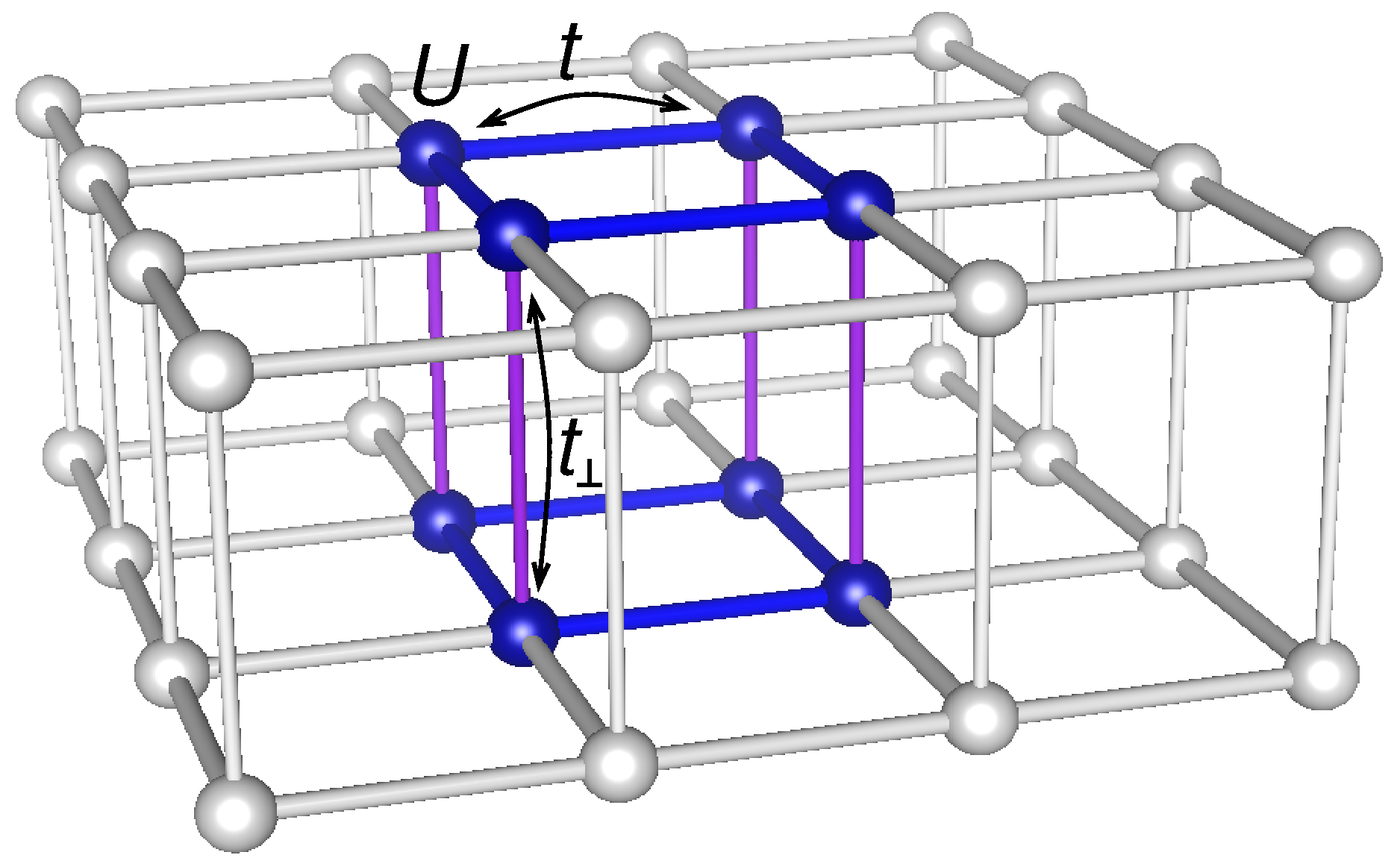}
\caption{(Color online) Cartoon of the bilayer Hubbard Hamiltonian. The $N_c=2 \times 4$ region studied here using DCA is shown in darker colors. }
\label{Fig1:model}
\end{figure}

In fact, the bilayer Hubbard model (see Fig.~\ref{Fig1:model}) was investigated by several groups
within the dynamical mean field theory approximation
(DMFT)~\cite{Fuhrmann2006,Hafermann2009} and cellular
DMFT~\cite{Kancharla2007}. Transitions from metal to band insulator at
small $U/t$ and from Mott to band insulator at large $U/t$ were
reported with increasing $t_{\perp}/t$.  However, by definition
spatial fluctuations are completely ignored in DMFT~\cite{Georges1996}
and such features like an intermediate non-Fermi-liquid phase in the
single-layered Hubbard model at half
filling~\cite{Zhang2007,Park2008,Gull2008} are not captured by this
approach.  Moreover, only a small cluster size of $N_c=2\times2$ (two
sites in one layer and two sites in another layer) was used in
previous cellular DMFT calculations for the bilayer Hubbard
model~\cite{Kancharla2007}.  For such sizes, the $C_4$ rotational
symmetry of the square lattice is broken, resulting in an artificially
enhanced local pair within each plane as the interaction $U/t$ is
switched on.  As reported in Ref.~\onlinecite{Lee2010}, such a choice
is unable to describe an intermediate orbital-selective phase in a
two-orbital Hubbard model. On the other hand, the dynamical cluster
approximation (DCA) with clusters sizes of $N_c=2 \times 16$
(Ref.~\onlinecite{Maier2011}), the determinant quantum Monte Carlo
method~\cite{Bouadim2008} and the Gutzwiller
approximation~\cite{Lanata2009} -which do not suffer from the previous
cluster limitations- were recently employed mainly to understand the
nature of the superconducting state in the bilayer Hubbard model away
from half filling.  In view of the above results, and due to the
limitations of the various approaches used, there is still not a full
understanding of the transition from band to Mott insulating behavior
as a function of $t_{\perp}/t$ in the bilayer Hubbard model at half
filling.

In the present work we concentrate on this question and study the
bilayer Hubbard model at and away from half filling as the band
splitting ratio $t_{\perp}/t$ is increased up to $4$ at a large
interaction strength of $U/t=6.0$ in the framework of the
DCA~\cite{Hettler1998,Maier2005}.  We consider one plaquette in each
layer, i.e., a cluster size of $N_c=2 \times 4$, which keeps the
rotation symmetry of the square lattice and allows for short-range
spatial fluctuations.  We focus on the lower temperature regime, which
couldn't be accessed in previous studies~\cite{Maier2011} due to the
larger cluster sizes. We use an interaction-expansion continuous-time
quantum Monte Carlo algorithm as an impurity
solver~\cite{Rubtsov2005,Assaad2007,Gull2011}. Note that the value of
the critical interaction strength for the metal-insulator transition
is sensitive to the level of approximation
considered~\cite{Fuhrmann2006,Hafermann2009,Kancharla2007,Tocchio2013}.
Within DCA and for the cluster sizes considered in this work, the
value $U/t=6.0$ is deep in the insulating phase.
In order to distinguish
between a Mott and a band insulating regime we analyze the behavior of (i) the density of states
near the Fermi level,
 (ii) the imaginary part of the self energy at the lowest Matsubara
frequency and (iii) the momentum resolved electron density.

At small interlayer hopping $t_{\perp}/t$ we observe Mott insulating
behavior in the DCA momentum sectors ($\pi,0$)/($0,\pi$) and band
insulating behavior in the DCA momentum sectors ($0,0$)/($\pi,\pi$) as
was also obtained in cellular DMFT with $N_c=4$ in the single-band
Hubbard model~\cite{Park2008}.  With increasing $t_{\perp}/t$ the Mott
behavior is rapidly suppressed in the DCA momentum sectors ($\pi,0$)
and ($0,\pi$), and at $t_{\perp}/t\simeq2.5$ it becomes dominant in
the DCA momentum sectors ($\pi,\pi$) of the bonding band and ($0,0$)
of the anti-bonding band. At $t_{\perp}/t\simeq3.0$ a tiny scattering
rate is observed in all momentum sectors at the Fermi level,
indicating a transition from a Mott to a band insulator. Such a
momentum-dependent evolution of the insulating state has not been
reported in previous studies and will be analyzed in detail in the
present work.

\begin{figure}
\includegraphics[width=0.48\textwidth]{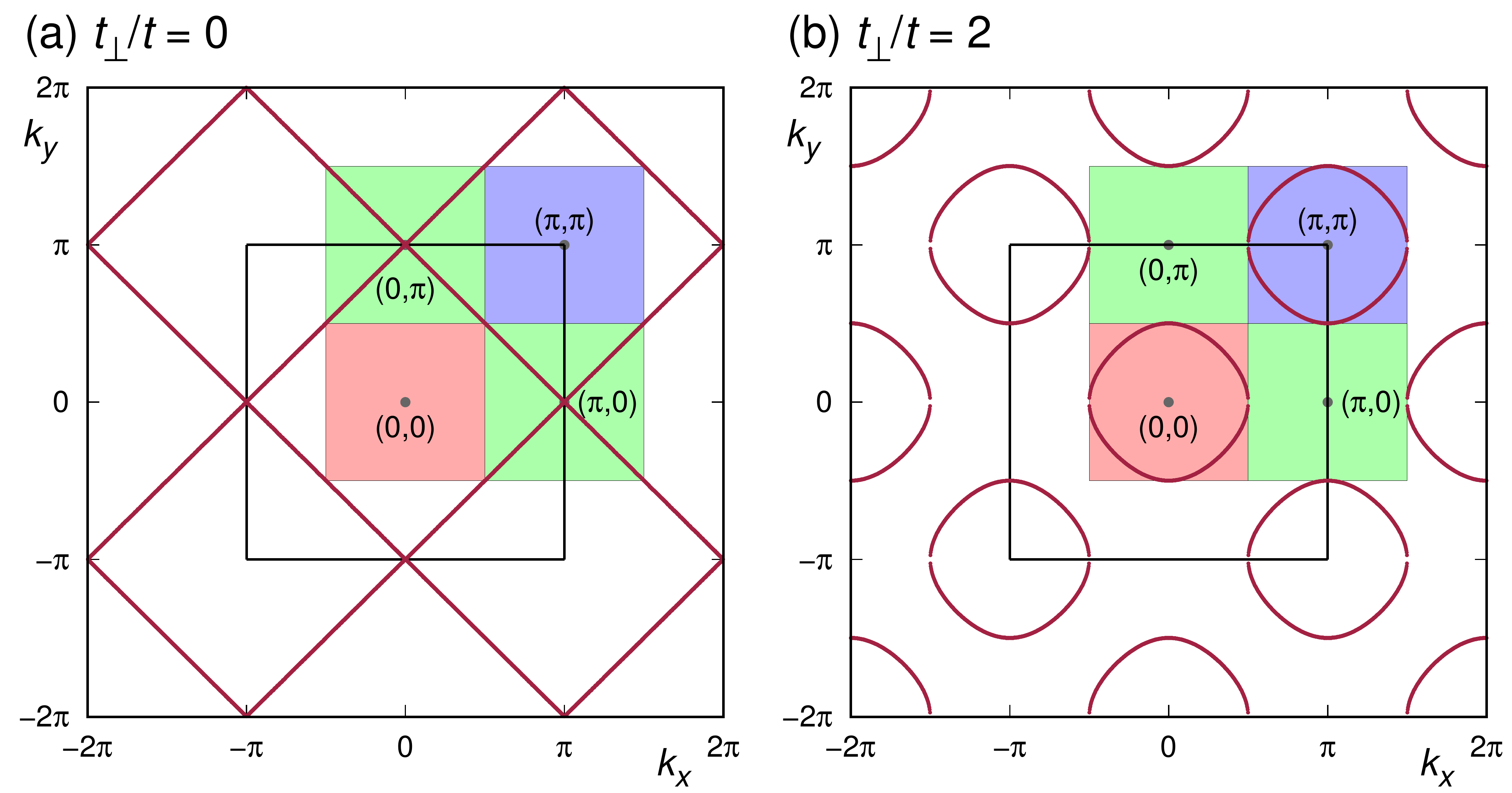}
\caption{(Color online) The Fermi surface for (a) $t_{\perp}/t=0.0$
  and (b) 2.0 at half-filling in the weak coupling limit of
  $U/t=0$. The colored areas indicate the patches for momentum
  clusters $\rm{K}=(0,0)$, $(\pi,0)$, $(0,\pi)$, and $(\pi,\pi)$ of
  the dynamical cluster approximation with $N_c = 2 \times
  4$.}\label{Fig1:sectors}
\end{figure}

In particular, we find a monolayer plaquette singlet Mott insulator
(m-PSMI) with strong intralayer plaquette order but weak interlayer
antiferromagnetic (AF) correlations at small $t_{\perp}/t$ and a
bilayer plaquette singlet Mott insulator (b-PSMI) with strong AF
correlations between plaquettes belonging
to different layers  at intermediate values
of $t_{\perp}/t$. At the critical value $t_{\perp}/t\simeq3.0$
we observe
a transition from a Mott to a band insulating state with tiny
intralayer spin-spin correlations. Such a momentum-dependent phase
behavior results from the competition and cooperation of short-range
spatial fluctuations and interlayer coupling $t_{\perp}$ with the help
of the Coulomb interaction $U$.

The paper is organized as follows. In Sec.~\ref{Formalism} we present
the model and dynamical cluster approximation. In Sec.~\ref{Results}
we present the density of states, the self-energy and electron density
at each DCA momentum sector as well as spin-spin correlations and we
discuss the nature of Mott insulator and band insulator at
half-filling as well as non-Fermi liquid behavior  away from half-
filling. Finally, in Sec.~\ref{Conclusions} we summarize our findings.

\section{Model and method}\label{Formalism}

The bilayer Hubbard Hamiltonian can be written as
\begin{eqnarray}
	H&=&-\sum_{\langle ij\rangle m\sigma}
	t_m(c^+_{jm\sigma}c_{im\sigma}+{\rm h.c.})-
        \mu\sum_{im\sigma}n_{im\sigma}\nonumber\\
	&-& t_\perp\sum_{i\sigma}(c^+_{i1\sigma}c_{i2\sigma}+{\rm
	h.c.})+U\sum_{im}n_{im\uparrow}n_{im\downarrow}
	\label{eq:1},
\end{eqnarray}
where $c_{im\sigma}(c^{\dag}_{im\sigma})$ annihilates (creates) an
electron with spin $\sigma$ at site $i$ and layer $m \in (1,2)$, and
$\mu$ is the chemical potential. $t_m$ is the intralayer hopping
matrix element between sites $i$ and $j$ in layer $m$ and $t_{\perp}$
is the interlayer hopping parameter which induces a band splitting
into a bonding and an anti-bonding band. For $t_m = t$ ($m$=1,2) with
$t$ as energy unit throughout this paper, the energy dispersion is
given as $\epsilon^{\text{A,B}} ({\rm k}) = \epsilon ({\rm k}) \pm
t_{\perp}$, where $\epsilon ({\rm k}) = -2t (\cos {\rm k_x} + \cos {\rm
  k_y})$ and 'A' and 'B' indices denote anti-bonding (antisymmetric)
($+$) and bonding (symmetric) ($-$) states,
respectively~\cite{Fuhrmann2006}.

The DCA is the cluster extension of single-site DMFT and the
self-consistent equation can be written in momentum space with the
assumption that the self-energy is constant in the Brillouin zone
sectors that are considered.  The cluster Green's functions are
calculated by integration of each sector:
\begin{eqnarray}\label{DCAE}
\overline{G}_{\sigma}({\bf K},i\omega_n)=\frac{1}{N}\sum_{\tilde{\bf K}}
\frac{1}{i\omega_n+\mu-\epsilon_{{\bf
K+\tilde{K}}}^{{\text{A,B}}}-\Sigma_{\sigma}({\bf K},i\omega_n)},
\end{eqnarray}
where $N$ is the number of ${\tilde{\bf K}}$ points in each Brillouin
zone sector, $\mu$ the chemical potential, ${\bf K}$ is the cluster
momentum, $\epsilon_{{\bf K+\tilde{K}}}^{{\text{A,B}}}$ is the
dispersion relation for antibonding and bonding states, $\omega_n$ are
the Fermionic Matsubara frequencies, and the summation over
$\tilde{\bf K}$ is performed in each Brillouin zone sector.  In our
calculations, we considered a DCA cluster with $N_c = 2\times 4$,
where ${\bf K} = (0,0)$, $(0,\pi)$, $(\pi,0)$ and $(\pi,\pi)$  (see Fig.~\ref{Fig1:sectors}).  The
converged self-energy $\Sigma_{\sigma}({\bf K},i\omega_n)$ is
evaluated by means of Eq.~(\ref{DCAE}) and the Dyson equation and we
employed the interaction expansion continuous-time quantum Monte Carlo
approach as an impurity
solver~\cite{Rubtsov2005,Assaad2007,Gull2011}. All calculations
presented below are for a temperature $T/t=0.1$ and more than $5
\times 10^6$ QMC samplings are employed to measure the impurity
Green's function.

In the non-interacting case ($U/t=0$), the model shows a band
insulating state at $t_{\perp}/t\geq4$ due to the complete separation
of bonding and anti-bonding bands, characterized by a formation of
fully localized dimers between layers at half-filling.

\section{Results}\label{Results}

\subsection{Half-filling}

\begin{figure}[htb]
\includegraphics[width=0.45\textwidth]{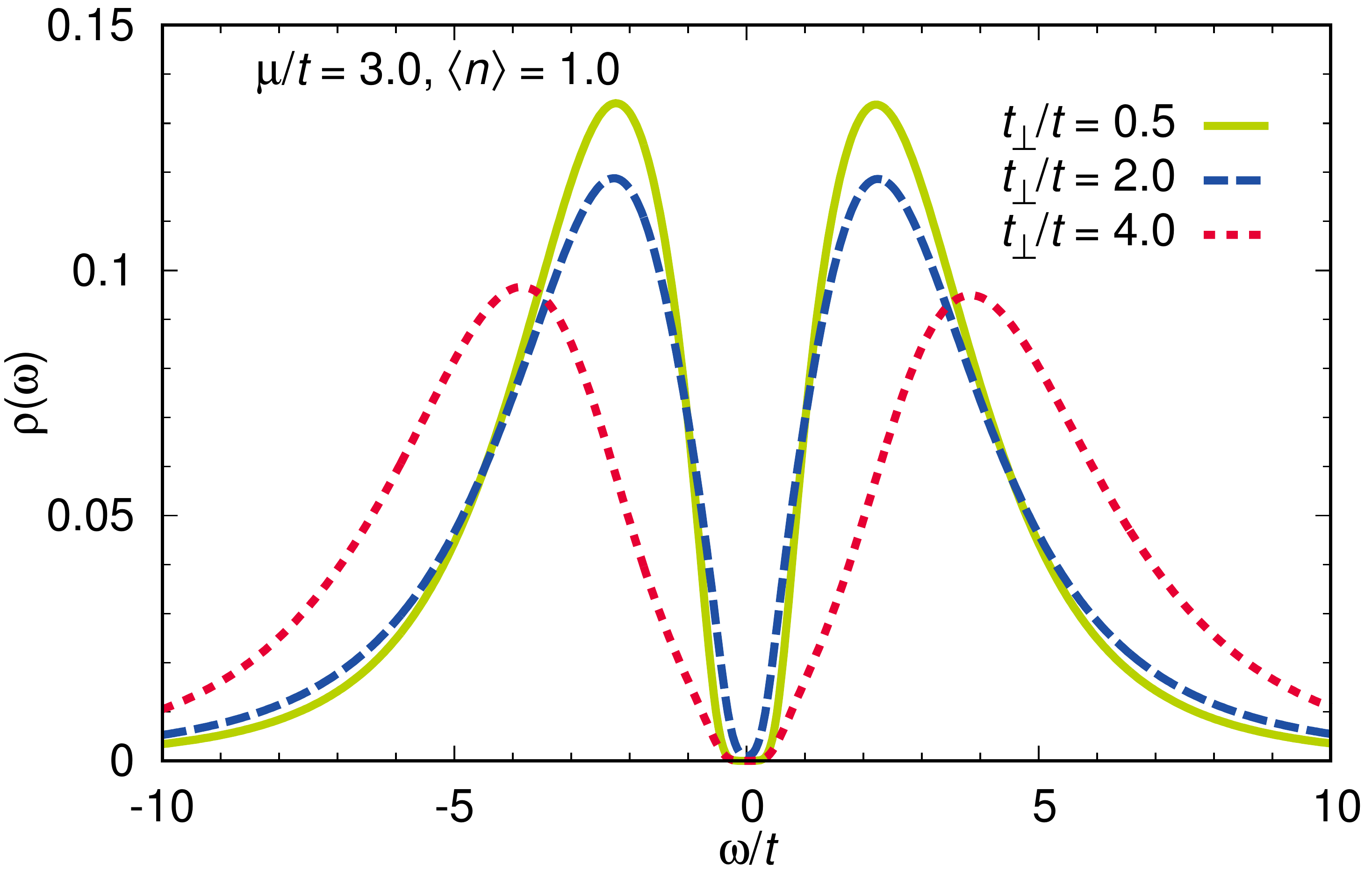}
\caption{(Color online) The density of states $\rho(\omega)$ for
  chemical potentials $\mu / t=3.0$ with half-filling at $U/t=6.0$ and
  $T/t=0.1$. The band splitting is $t_{\perp}/t=0.5$ (solid line),
  $2.0$ (dashed line), and $4.0$ (dotted line). The Fermi level is at
  $\omega=0$.}\label{Fig1:DOS}
\end{figure}

In Fig.~\ref{Fig1:DOS} we show the density of states $\rho(\omega)$ of
the model at $T/t=0.1$ and $U/t=6.0$ for small, intermediate, and
large band splittings with ratios $t_{\perp} / t=0.5$, $2.0$, and
$4.0$, respectively.  At $t_{\perp} / t = 0.5$ the gap is nonzero and
dominated by the interaction $U$, and the bonding and antibonding
states both contribute to the lower and upper Hubbard bands. On the
other hand, at $t_{\perp}/t=4.0$, the gap, which is also significant,
is dominated by the band splitting $2t_{\perp}$. In this case, only
the bonding state contributes to the band below the Fermi level while
the antibonding one contributes to the band above the Fermi level. At
intermediate $t_{\perp} / t=2.0$ the gap amplitude is reduced to a
smaller value compared to the gap amplitudes at $t_{\perp} / t=0.5$
and $4.0$. Here, both bonding and antibonding states significantly
contribute to the bands above and below the Fermi level. However, the
peak positions remain located at the same frequency as those for
$t_{\perp} / t=0.5$ which indicates that Mott behavior is still
present, i.e.  the peak position is only dependent on $U/t$ and not on
$t_{\perp} / t$.

\begin{figure}[htb]
\includegraphics[width=0.45\textwidth]{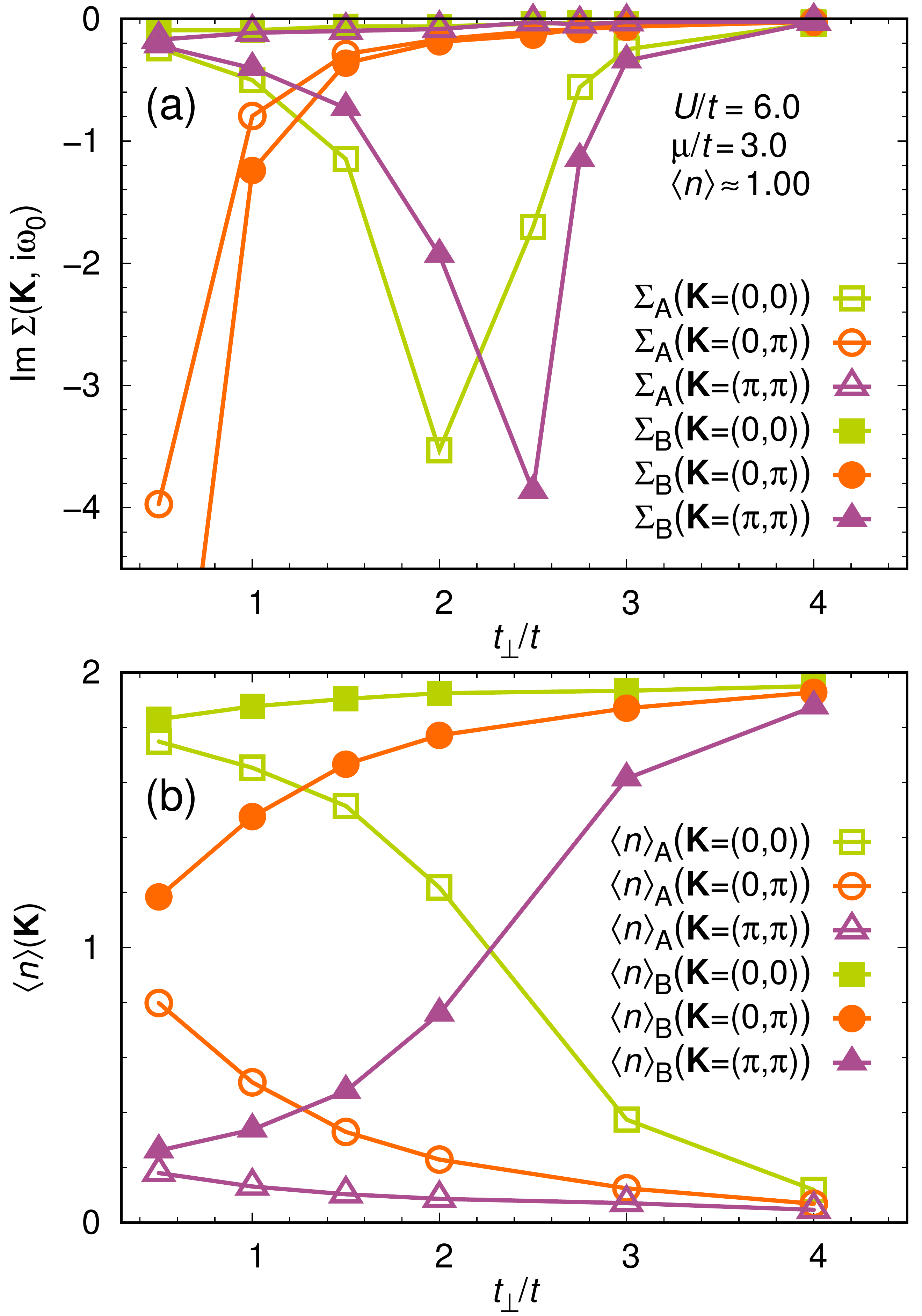}
\caption{(Color online) (a) Imaginary part of the DCA cluster
  self-energy ${\rm Im} (\Sigma({\bf K},\omega_0))$ at the lowest
  Matsubara frequency $\omega_0$ and (b) electron density $\langle
  n({\rm K})\rangle$ as a function of $t_{\perp}/t$ at each DCA
  momentum sector ${\rm K}$.  Results are shown at a temperature
  $T/t=0.1$ and an interaction strength $U/t=6.0$ for chemical
  potential $\mu/t=3.0$ (half-filling $\langle n_i\rangle =
  1.00$). ${\bf K}$  are the DCA cluster momenta and indices
  ${\rm A}$ and ${\rm B}$ stand for antibonding and bonding,
  respectively.}\label{Fig2:momentum}
\end{figure}

In order to obtain a deeper insight into the nature of the phases as a
function of the band splitting, we analyze in the following the
cluster self-energy at the various DCA momentum sectors $\bf K$.
Figs.~\ref{Fig2:momentum} (a) and (b) show the imaginary part of the
DCA cluster self-energy ${\rm Im}\, \Sigma({\bf K},\omega_0)$ at the
lowest Matsubara frequency $\omega_0$ and the electron density
$\langle n({\bf K})\rangle$, respectively, as a function of
$t_{\perp}/t$ for $U/t=6.0$ at half filling where $\mu / t=3.0$. Both
quantities are shown for bonding (B) and antibonding (A) bands for the
cluster momenta ${\bf K}=(0,0)$, $ (0,\pi)$ (identical to $(\pi,0)$)
and $(\pi,\pi)$

\begin{figure}[htbp]
\centering
\includegraphics[width=0.45\textwidth]{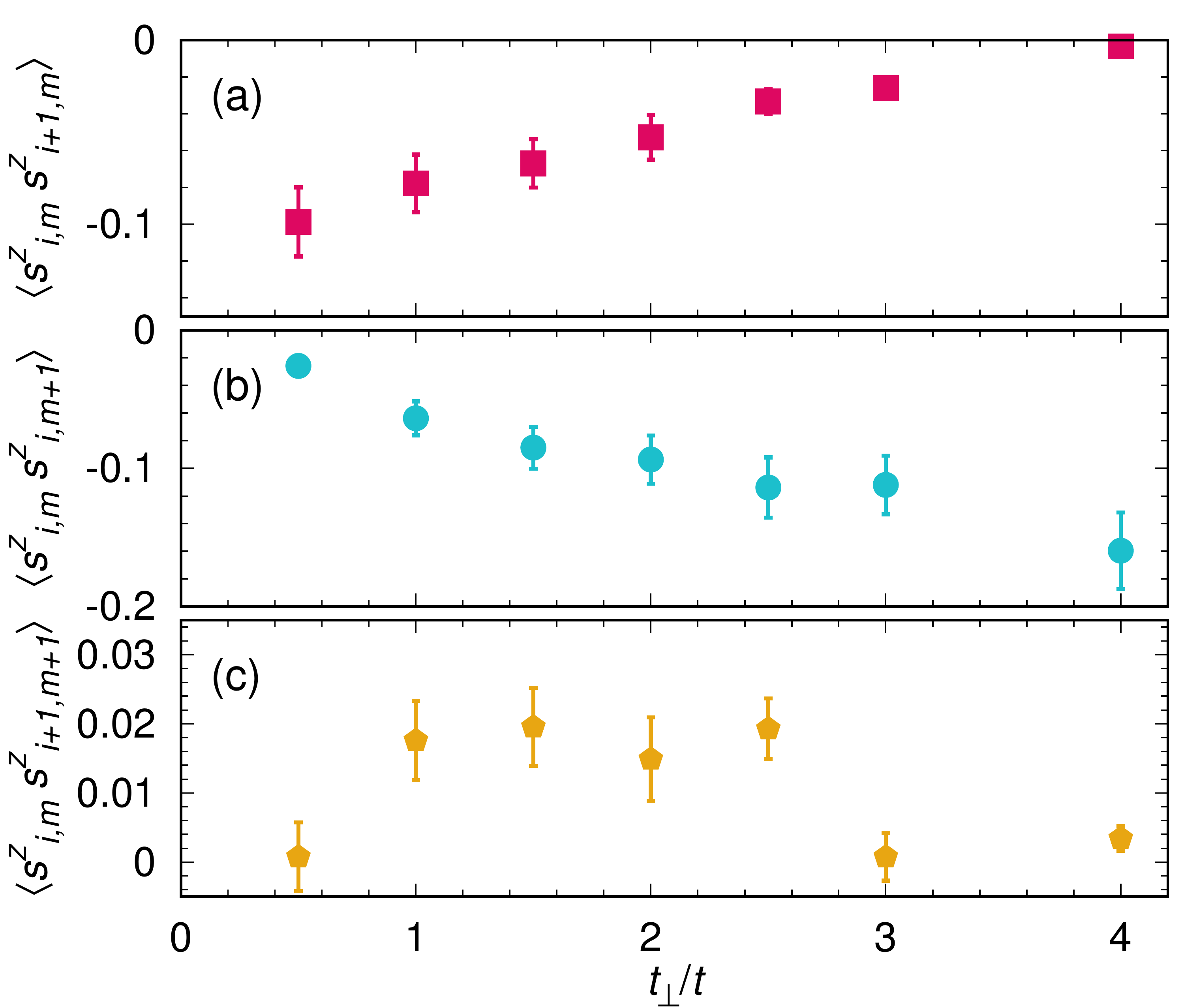}
\caption{(Color online) The spin-spin correlations $\langle s_{i,m}^z
  s_{i',m'}^z\rangle$ as a function of $t_{\perp}/t$ (a) between $i$ and
  nearest neighbors site $i'=i+1$ in the same layer $m$, (b) between
  layer $m$ and $m'=m+1$ in the same site $i$, and (c) between $i$ and
  nearest neighbor site $i'=i+1$ as well as layer $m$ and $m'=m+1$ for
  $U/t=6.0$ and $T/t=0.1$.}
\label{Sup1}
\end{figure}

At small band splitting $t_{\perp} / t < 0.5$,  momentum selective
Mott insulating phases with large scattering rates ${\rm
  Im}\,\Sigma({\bf K},\omega_0)$ are present in both bonding and
antibonding bands at ${\bf K}=(\pi,0)$ and $(0,\pi)$ momentum sectors
where the electron density $\langle n({\bf K})\rangle$ indicates almost
half-filling, while the scattering rates at ${\bf K}=(0,0)$ and
$(\pi,\pi)$ momentum sectors remain small suggesting a band insulating
behavior with empty or fully filled electron density $\langle n({\bf
  K})\rangle$ in these sectors.  Further analysis on spin-spin
correlations (Fig.~\ref{Sup1}) shows strong intralayer but weak
interlayer antiferromagnetic correlations (Fig.~\ref{Sup1} (a) and
(b)).  Moreover, the spin-spin correlation $\langle s_{i,m}^z
s_{i+1,m'}^z \rangle$ for different layers and nearest neighbor sites
displays very weak ferromagnetic order at $t_{\perp} / t = 0.5$ (see
Fig.~\ref{Sup1} (c)).  We denote the state at small $t_{\perp} / t$
m-PSMI (monolayer plaquette singlet Mott insulator).  This m-PSMI
state has been reported to be present in the one-band Hubbard model
investigated by a 4-site cluster-DMFT approach~\cite{Gull2008}.

As $t_{\perp} / t$ increases, the scattering rates ${\rm
  Im}\,\Sigma({\bf K},\omega_0)$ at the ${\bf K}=(\pi,0)$ and
$(0,\pi)$ momentum sectors rapidly decrease towards zero while the
scattering rate at the $(\pi,\pi)$ momentum sector of the bonding band
(see $\Sigma_B({\bf{K}}=(\pi,\pi))$ in Fig.~\ref{Fig2:momentum} (a))
and that at the $(0,0)$ momentum sector of the antibonding band (see
$\Sigma_A({\bf{K}}=(0,0))$ in Fig.~\ref{Fig2:momentum} (a)) develop
dramatically.  In terms of the electron density $\langle n({\bf
  K})\rangle$ in Fig.~\ref{Fig2:momentum} (b), the $(\pi,\pi)$
momentum sector of the bonding band and the $(0,0)$ momentum sector of
the antibonding band are filled with about one electron at $t_{\perp} /
t=2.2$. The strong scattering is caused by interactions between
electrons in the $(\pi,\pi)$ momentum sector of the bonding band and
the $(0,0)$ momentum sector of the antibonding band at $t_{\perp} /
t=2.2$.  The spin-spin correlations exhibit intermediate intralayer as
well as interlayer antiferromagnetic correlations. Specifically, the
ferromagnetic correlations for different layers and nearest neighbor
sites reach a maximum in Fig.~\ref{Sup1} (c).  This means that the
plaquette singlet orderings in each layer develop AF correlations. We
denote this state bilayer plaquette singlet Mott insulator (b-PSMI).

When $t_{\perp}/t$ is further increased, tiny scattering rates
${\rm Im}\,\Sigma({\bf K},\omega_0)$ are observed in the $(\pi,\pi)$
momentum sector of the bonding band and the $(0,0)$ momentum sector of
the antibonding band between $t_{\perp}/t = 3.0$ and $4.0$, while the
electron densities $\langle n({\bf K})\rangle$ are not fully occupied
(or empty) as shown in Figs.~\ref{Fig2:momentum}~(a) and (b).  Such an
insulator has been denominated a covalent band
insulator~\cite{Sentef2009}.

In the region $t_{\perp} / t > 4.0$, all scattering rates disappear
and the system is in a band insulating state, where the electron
density $\langle n({\bf K})\rangle$ are fully filled (or empty) in all
momentum sectors and all bands. The ratio $t_{\perp} / t $ at which
the Mott to band insulator phase transition happens is smaller than
the the ratio for the metal-insulator transition in the noninteracting
case ($t_{\perp} / t = 4.0$). This is due to the fact that strong
correlation narrows the bandwidth of the bonding and antibonding bands
and consequently a smaller band splitting is required for opening a
band gap. The interlayer spin-spin correlations indicate a strong
dimer state, while the remaining spin-spin correlations are extremely
weak as shown in Fig.~\ref{Sup1}.  The state at $t_{\perp} / t = 4.0$
is a band insulator with isolated dimers between layers.

The momentum-dependent evolution of the insulating states can be well
understood from the evolution of the Fermi surface in the weak
coupling limit as a function of $t_{\perp}$. As shown in
Fig.~\ref{Fig1:sectors} (a), at $t_{\perp}/t=0$, all the Fermi
surfaces are located in the momentum sectors $(\pi,0)$ and $(0,\pi)$,
indicating that poles determined by $\omega+\mu-\epsilon_{{\bf
    K+\tilde{K}}}^{{\text{A,B}}}-{\rm Re} \Sigma_{\sigma}({\bf
  K},\omega)|_{\omega=0}=0$ are only present in these two sectors. As
the interaction $U$ becomes larger than the critical value of the Mott
metal-to-insulator transition, the opening of a gap at the Fermi level
indicates ${\rm Im} G({\bf K},\omega=0)\rightarrow0$, which requires
large scattering rates at the positions of the poles.  Therefore, Mott
physics occurs in momentum sectors $(\pi,0)$ and $(0,\pi)$.  On the
other hand, at $t_{\perp}/t=2.0$ (see Fig.~\ref{Fig1:sectors} (b)),
almost all the Fermi surfaces from the bonding band enter the momentum
sector $(\pi,\pi)$ while those from anti-bonding band are mostly in
the momentum sector $(0,0)$.  In order to be a Mott insulator at large
interaction $U$, large scattering rates are again inevitable in the
momentum sectors $(\pi,\pi)$ of the bonding band and $(0,0)$ of the
anti-bonding band since poles are now located in these sectors. Thus,
the momentum-dependent evolution of the insulating states is a
consequence of the evolution of the Fermi surface.

\subsection{Away from half-filling}

\begin{figure}
\includegraphics[width=0.45\textwidth]{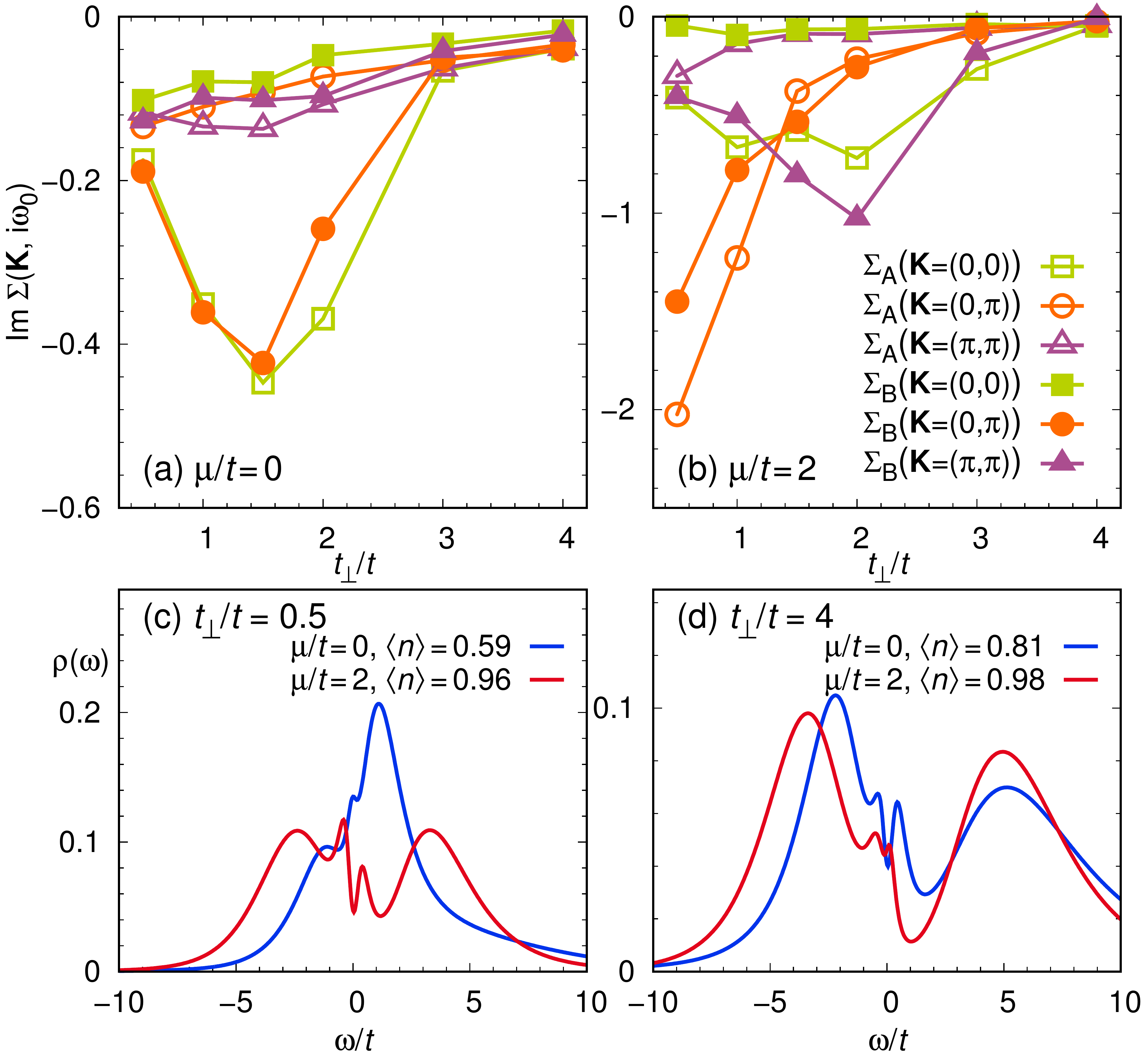}
\caption{(Color online) (a), (b) Imaginary part of the DCA cluster
  self-energy ${\rm Im} (\Sigma({\bf K},\omega_0))$ at the lowest
  Matsubara frequency $\omega_0$ as a function of
  $t_{\perp}/t$. Results are shown at a temperature $T/t=0.1$ and an
  interaction strength $U/t=6.0$ for different values of the chemical
  potential: (a) $\mu/t=0.0$ (heavily doped case) and (b) 2.0 (close
  to half-filled case). The indices '${\bf K}$' represent the DCA
  cluster momenta and '${\rm A}$' and '${\rm B}$' indicate antibonding
  and bonding, respectively. (c), (d) Density of states $\rho(\omega)$
  for different chemical potentials $\mu / t$ at $U/t=6.0$ and
  $T/t=0.1$. The band splitting is (c) $t_{\perp}/t=0.5$ and (d)
  $t_{\perp}/t=4.0$. The Fermi level is at
  $\omega=0$.}\label{Fig3:momentum}
\end{figure}

We concentrate in what follows on the origin of non-Fermi liquid or
pseudogap, which has been discussed extensively in the
literature~\cite{Stanescu2006,Sakai2009,Gull2010,Sordi2011,Sakai2012},
based on the bilayer Hubbard model away from half-filling.  In the
heavily doped case ($\mu / t=0.0$), we observe at $t_{\perp}/t = 0.5$
for $T/t=0.1$ a Fermi liquid like metallic behavior with ${\rm
  Im}\,\Sigma({\bf K}, \omega_0)$ approaching small finite values due
to the finite temperature effect (see Fig.~\ref{Fig3:momentum} (a)). A
quasiparticle peak is present at the Fermi level in the DOS (see
Fig.~\ref{Fig3:momentum} (c)).  Close to the half-filled case, such as
$\mu / t=2.0$ at $t_{\perp}/t = 0.5$, large but finite scattering
rates are observed in the antibonding ${\bf K}=(0,\pi)$ and the
bonding ${\bf K}=(0,\pi)$ sectors due to the enhancement of intralayer
short-range AF correlations. As a result, a pseudogap appears in the
DOS at the Fermi level (see Fig.~\ref{Fig3:momentum} (c)), reminiscent
of the non-Fermi liquid behavior observed in the single-band and
multi-band Hubbard
models~\cite{Zhang2007,Park2008,Lee2010,Gull2008,Lee2010,Lee2011}. At
$t_{\perp}/t =4.0$, though the scattering rates in all momentum
sectors and all bands vanish due to the strong interlayer AF
correlations (see Fig.~\ref{Fig3:momentum} (b)), pseudogaps exist in
both heavily doped and nearly half-filled cases, indicating that a
strong scattering rate is not a necessary condition for the appearance
of non-Fermi liquid behavior; rather the short-range AF correlations
alone can be responsible for the non-Fermi liquid behavior.

\begin{figure}[htbp]
\includegraphics[width=0.45\textwidth]{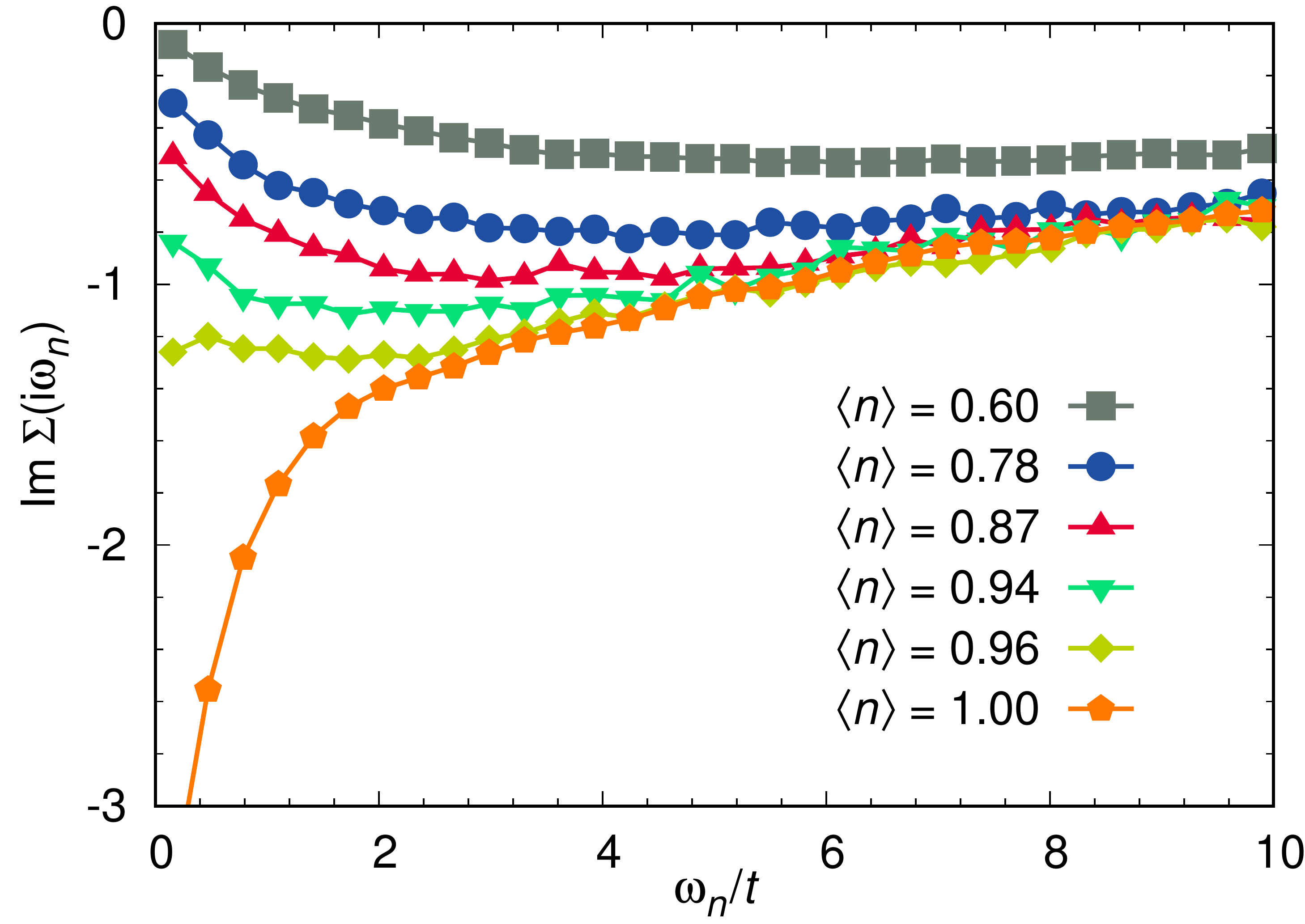}
\caption{(Color online) The imaginary part of on-site self-energy
  $\rm{Im}\,\Sigma ({i\omega_n})$ as a function of Matsubara frequency
  $\omega_n$ for $T/t=0.05$, $t_{\perp}/t=0.5$, and $U/t=6.0$ in
  different electron densities $\langle n\rangle$.}
\label{Fig6:selfenergy}
\end{figure}

Furthermore, in order to confirm the transitions from the Fermi liquid
to Mott insulator via non-Fermi liquid as a function of electron
doping (or chemical potential), we plot the ${\rm
  Im}\,\Sigma(i\omega_n)$ at the lower temperature $T/t=0.05$ and
$t_{\perp}/t = 0.5$ for $U/t=6.0$ in Fig.~\ref{Fig6:selfenergy}.  In
the case of $\langle n\rangle=0.60$, $\rm{Im}\,\Sigma (i\omega_0)$
converges to almost zero indicating a Fermi-liquid behavior. As the
electron density $\langle n\rangle$ approaches half-filling between
$\langle n\rangle=0.87$ and $0.96$, $\rm{Im}\,\Sigma (i\omega_0)$
converges to finite values. Such a behavior is typical for non-Fermi
liquid states.  Finally, at half-filling with $\langle n\rangle=1.00$
the system is a Mott insulator with $\rm{Im} \Sigma (i\omega_0)$
exhibiting a diverging behavior.

\section{Conclusions}\label{Conclusions}

In conclusion, we investigated a two-dimensional bilayer Hubbard model
on the square lattice as a function of band splitting and doping at a
large interaction value $U/t=6.0$ by means of the dynamical cluster
approximation with a $N_c=2 \times 4$ site cluster with short-range
spatial as well as quantum fluctuations.  The scattering rate, which
indicates the degree of Mott physics, rapidly disappears in the momentum
sectors ($\pi,0$) and ($0, \pi$) with increasing interlayer hopping
$t_{\perp}/t$.  In fact we find
a momentum-selective phase reentrant behavior from band insulating
states
at weak interlayer hopping $t_{\perp}/t\simeq0.5$ to Mott insulating
behavior
at $t_{\perp}/t\simeq2.0$ and then from Mott insulating to band
insulating behavior at strong interlayer hopping $t_{\perp}/t\simeq
3.0$.  These transitions are identified from the scattering rates in
the ($\pi,\pi$) momentum sector of the bonding band and the ($0,0$)
momentum sector of the anti-bonding band at half filling.  Interesting
phases are established with two consecutive phase transitions from a
monolayer plaquette singlet Mott insulator (m-PSMI) to a band
insulator through an intermediate phase, called bilayer plaquette
singlet Mott insulator (b-PSMI), where Mott physics is more present in
the $(\pi,\pi)$ sector of the bonding band and the $(0,0)$ of the
anti-bonding band, rather than in the $(\pi,0)$ and $(0, \pi)$ sectors
as usually observed in an antiferromagnetic Mott insulator.  We
attribute the unusual consecutive phases to competition and
cooperation between short-range spatial correlations with quantum
fluctuations and interlayer hopping $t_{\perp}/t$.  The
transition of Mott to band insulator with the absence of large
scattering rates in all momentum sectors is found at large interlayer
hopping $t_{\perp}/t\simeq3.0$.
Furthermore, since the momentum-dependent
evolution of the insulating behavior with $t_{\perp}/t$  is strongly controlled by the
evolution of the Fermi surface in the weak-coupling limit,
we expect that consideration of larger clusters and different geometries
in DCA won't change this scenario qualitatively.


Finally we also find that away from half
filling, non-Fermi liquid behavior is dominated by antiferromagnetic
correlations rather than the finite scattering rate at the Fermi
level. We suggest that this non-Fermi liquid behavior might be related
to anomalous phenomena like the Fermi arc or hole pocket. We expect
that momentum-selective phenomena may exist in many cases which calls
for further studies in various models and real materials where the
short-range spatial fluctuations are emphasized.

\section{Acknowledgement}\label{Acknowledgement}

We would like to thank J. Kune\v{s}, L. F. Tocchio, F. Becca, I.I. Mazin, M. Capone,
and C.  Gros for useful discussion. We gratefully acknowledge
financial support from the Deutsche Forschungsgemeinschaft through
grants FOR 1346 and SFB/TRR 49. Y.Z is supported by National Natural
Science Foundation of China (No. 11174219), Shanghai Pujiang Program
(No.  11PJ1409900), Research Fund for the Doctoral Program of Higher
Education of China (No.  20110072110044) and the Program for Professor
of Special Appointment (Eastern Scholar) at Shanghai Institutions of
Higher Learning. H.L is supported by ERC/FP7 through the Starting Independent Grant "SUPERBAD", Grant Agreement No. 240524.

\end{document}